\documentclass[aps,prl,twocolumn,superscriptaddress]{revtex4-2}  
\usepackage{amsfonts}
\usepackage{bbding}
\usepackage{amssymb}
\usepackage[title]{appendix}
\usepackage{color}
\usepackage{amsmath}
\usepackage{graphicx}
\usepackage{subfigure}
\usepackage{txfonts}
\usepackage{bm}
\usepackage[normalem]{ulem}
\usepackage{natbib}
\usepackage{calrsfs}
\usepackage{tikz}
\usetikzlibrary{graphs, positioning, quotes} 
\usetikzlibrary{matrix}
\usepackage[colorlinks,linkcolor=blue,citecolor=blue,urlcolor=blue]{hyperref}
\usepackage{orcidlink}
\preprint{APS/123-QED}

\DeclareMathAlphabet{\pazocal}{OMS}{zplm}{m}{n}
\newcommand{\Ls}{\pazocal{L}}

\newcommand{\Ds}{\pazocal{D}}

\newcommand{\blueit}[1]{\textcolor{blue}{\textit{#1}}}

\begin{document}

\title{Symmetry-Resolved Criticality in Dissipative Quantum Phase Transitions}

\author{Xingli Li\orcidlink{0000-0003-2339-6557}}
\affiliation{Department of Physics, The Chinese University of Hong Kong, Shatin, New Territories, Hong Kong, China}
\affiliation{The Chinese University of Hong Kong Shenzhen Research Institute, Shenzhen 518057, China}
\affiliation{Lanzhou Center for Theoretical Physics, Key Laboratory of Theoretical Physics of Gansu Province, Key Laboratory of Quantum Theory and Applications of MoE, Gansu Provincial Research Center for Basic Disciplines of Quantum Physics, Lanzhou University, Lanzhou 730000, China}

\author{Yan Li\orcidlink{0009-0007-7218-133X}}
\email{yanliphy@snnu.edu.cn}
\affiliation{School of Physics and Information Technology, Shaanxi Normal University, Xi’an, 710062, P. R. China}

\author{Jinghui Pi\orcidlink{0000-0002-6899-8132}}
\affiliation{The Chinese University of Hong Kong Shenzhen Research Institute, Shenzhen 518057, China}

\author{Wenlin Li\orcidlink{0000-0003-2032-0530}}
\affiliation{College of Sciences, Northeastern University, Shenyang 110819, China}
\affiliation{Research Center for Quantum Physics, Northeastern University, Shenyang 110819, China}

\author{\mbox{Shuai Geng}\orcidlink{0009-0000-8687-0493}}
\affiliation{School of Physics, Dalian University of Technology, Dalian 116024, China}

\author{Lin Shang\orcidlink{0009-0007-5840-2351}}
\affiliation{School of Physics, Dalian University of Technology, Dalian 116024, China}

\author{Jiasen Jin\orcidlink{0000-0002-7118-7488}}
\email{jsjin@dlut.edu.cn}
\affiliation{School of Physics, Dalian University of Technology, Dalian 116024, China}

\author{Yangqian Yan\orcidlink{0000-0002-3237-5945}}
\email{yqyan@cuhk.edu.hk}
\affiliation{Department of Physics, The Chinese University of Hong Kong, Shatin, New Territories, Hong Kong, China}
\affiliation{State Key Laboratory of Quantum Information Technologies and Materials, The Chinese University of Hong Kong, Hong Kong SAR, China}
\affiliation{The Chinese University of Hong Kong Shenzhen Research Institute, Shenzhen 518057, China}

\begin{abstract}
Symmetry breaking in mixed quantum states exhibits a rich strong-to-weak hierarchy, yet the use of symmetry as an operational probe of dissipative criticality remains largely unexplored.
In this work, we develop a symmetry-based framework that connects semiclassical phase space, trajectory-level dynamics, and ensemble responses, and demonstrate it in the two-dimensional dissipative XYZ model. 
We show that symmetry breaking manifests semiclassically as a bifurcation of the Husimi-$Q$ distribution and dynamically as a sharp change in a trajectory-level symmetry residual. 
At the ensemble level, the weak-symmetry sensitivity to a perturbed symmetry operator exhibits pronounced finite-size scaling, yielding an estimate of the dissipative critical point. 
Finally, we propose a trajectory-based quantum-circuit algorithm in which mid-circuit measurements condition quantum jumps and allow the proposed indicators to be extracted from circuit observables. 
Our work provides a symmetry-centered route to diagnosing dissipative phase transitions and critical scaling in open quantum many-body systems.
\end{abstract}
\date{\today}

\maketitle

\blueit{Introduction.--} Symmetry is a cornerstone of quantum phase transitions. 
While spontaneous symmetry breaking traditionally characterizes long-range order and critical behavior in closed many-body systems, extending this description to mixed states reveals a richer hierarchy~\cite{Sala2024,Ma2025f,Lessa2025}.
A density matrix may preserve a global weak symmetry, even as its underlying pure-state decompositions or correlation sectors manifest strong symmetry breaking~\cite{Kuno2024,Liu2025j,Guo2025a,Guo2025h,Ding2026,Zhang2025v,Sun2025c}.
This strong-to-weak hierarchy implies that conventional order parameters may fail to distinguish mixed-state phases, calling for diagnostics beyond pure-state criteria.

Open quantum many-body systems provide a natural dynamical setting for this hierarchy.
Their steady states are nonequilibrium mixed states governed by Liouvillian dynamics. 
These dynamics can also be unraveled into pure-state quantum trajectories ~\cite{Breuer2010,Dalibard1992,Molmer1993}.
Although Liouvillian symmetries are known to constrain steady states, symmetry has rarely been promoted from a structural constraint to an operational probe of scaling laws and critical points.
Existing diagnostics rely instead on conventional order parameters~\cite{Liu2025d,Wu2023b,Singh2022,Jin2018a,Maghrebi2016,Carollo2019,Savona2017,Jin2021}, Liouvillian spectra~\cite{Minganti2018,Huybrechts2020,Beaulieu2025,Rota2018a,Haga2024,Zhang2021d,Vicentini2018}, or mean-field analysis~\cite{Lee2013,Jin2016,Rose2016,Shang2026,Ye2021,Jo2019}, leaving the underlying symmetry structure largely detached from the determination of dissipative critical points and scaling behavior.
Consequently, a unified framework that integrates trajectory-level symmetry breaking and ensemble weak-symmetry sensitivity to directly uncover these critical properties is still lacking.

In this Letter, we address these gaps by developing a symmetry-breaking framework and demonstrating it in the two-dimensional dissipative spin-$1/2$ XYZ model on a square lattice~\cite{Lee2013}.
By combining phase-space analysis, quantum trajectories, and ensemble steady-state sensitivity, we identify complementary signatures of the dissipative transition.
Our key findings are as follows.
First, the steady-state Husimi-$Q$ distribution changes from a single peak to two symmetry-related peaks as the anisotropic interaction increases, signaling discrete symmetry breaking at the semiclassical level.
Second, we introduce a trajectory-level symmetry residual that quantifies the departure of a pure state from a symmetry sector.
This residual changes sharply near the transition region, revealing the onset of trajectory-level symmetry breaking.
Third, we probe the steady-state density matrix with a perturbed symmetry operator and identify signatures of criticality through the weak-symmetry sensitivity of the steady state.
The sensitivity is significantly enhanced in the same region, and its peak position follows finite-size scaling, providing an estimate of the critical point.

We propose a trajectory-based quantum-circuit (QC) algorithm for Lindblad dynamics, in which mid-circuit measurement outcomes determine whether quantum jumps occur and allow the symmetry residual to be extracted from circuit observables~\cite{Sander2025a,Endo2020,Yu2025e}.
Taken together, these results synthesize semiclassical phase-space signatures, trajectory-level symmetry breaking, and ensemble sensitivity into a coherent framework, suggesting a symmetry-centered route to diagnosing dissipative criticality in open quantum many-body systems.

\blueit{Model and Symmetry.--}We consider a dissipative spin-$1/2$ XYZ model with open boundary conditions, described by the Hamiltonian
$H=\sum_{\langle j,k\rangle}
\left(
J_x\sigma_j^x\sigma_k^x+
J_y\sigma_j^y\sigma_k^y+
J_z\sigma_j^z\sigma_k^z
\right)$, where $\sigma_j^\alpha$ are Pauli matrices of the $j$th spin and $\langle j, k\rangle$ denotes nearest-neighbor pairs ($\hbar=1$ hereinafter).
Under local spin decay, the system evolves according to the Lindblad master equation
\begin{equation}
\frac{d\rho}{dt}=\Ls\rho
=-i[H,\rho]
+\gamma\sum_{j=1}^{N}
\left(
\sigma_j^-\rho\sigma_j^+
-\frac{1}{2}\{\sigma_j^+\sigma_j^-,\rho\}
\right),
\label{eq:lindblad}
\end{equation}
where $\sigma_j^\pm=(\sigma_j^x\pm i\sigma_j^y)/2$ are the local spin raising and lowering operators, $\Ls$ is the Liouville superoperator, $N$ is the total number of spins, and the decay rate is set to $\gamma=1$.
This model is known to host a dissipative phase transition between a paramagnetic phase and a ferromagnetic phase with spontaneously broken
$\mathbb{Z}_2$ symmetry~\cite{Lee2013,Jin2016,Rota2017,Rota2018a,Biella2018,Li2022c,Verstraelen2023,Matsumoto2025a,Huybrechts2019}.
The symmetry operator is given by~\cite{Melo2025a}
\begin{equation} 
U=\prod_{j=1}^{N}\sigma_j^z .
\label{eq:z2_operator}
\end{equation}
Under this operation, $\sigma_j^{x,y}\rightarrow-\sigma_j^{x,y}$ and $\sigma_j^{\pm}\rightarrow-\sigma_j^{\pm}$.
The Hamiltonian is invariant under this transformation, while the jump operators acquire only a phase; hence the Lindblad generator is covariant under the global $\mathbb{Z}_2$ transformation,
\begin{equation}
\Ls\left(U\rho U^\dagger\right)
= U(\Ls\rho)U^\dagger .
\label{eq:weak_lindblad_symmetry}
\end{equation}
We refer to this covariance as the weak symmetry, because it constrains the ensemble density matrix rather than requiring each conditioned trajectory to remain in a definite symmetry sector.
At the trajectory level, we define the strong symmetry as imposing a more restrictive condition: the pure conditioned state must satisfy $U\rho=e^{i\varphi}\rho$.
Since symmetry breaking across a dissipative phase transition leaves signatures at both levels, it naturally motivates a symmetry-resolved characterization of the critical point and the nature of the transition.

\blueit{Phase-space signatures.--} We first seek semiclassical signatures of symmetry breaking from the Husimi-$Q$ distribution of steady states, which is defined as~\cite{Arecchi1972}
\begin{equation}
Q(\theta,\phi)
=
\frac{N+1}{4\pi}
\langle\theta,\phi|\rho_{\rm ss}|\theta,\phi\rangle ,
\label{eq:husimi}
\end{equation}
where $|\theta,\phi\rangle$ is a spin-coherent product state parameterized by the polar and azimuthal angles $\theta$ and $\phi$ on the collective Bloch sphere.
By resolving the steady-state weight over semiclassical spin configurations, this representation visualizes the symmetry structure of steady states in phase space.

\begin{figure}[!htpb]
    \centering
    \includegraphics[width=0.48\textwidth]{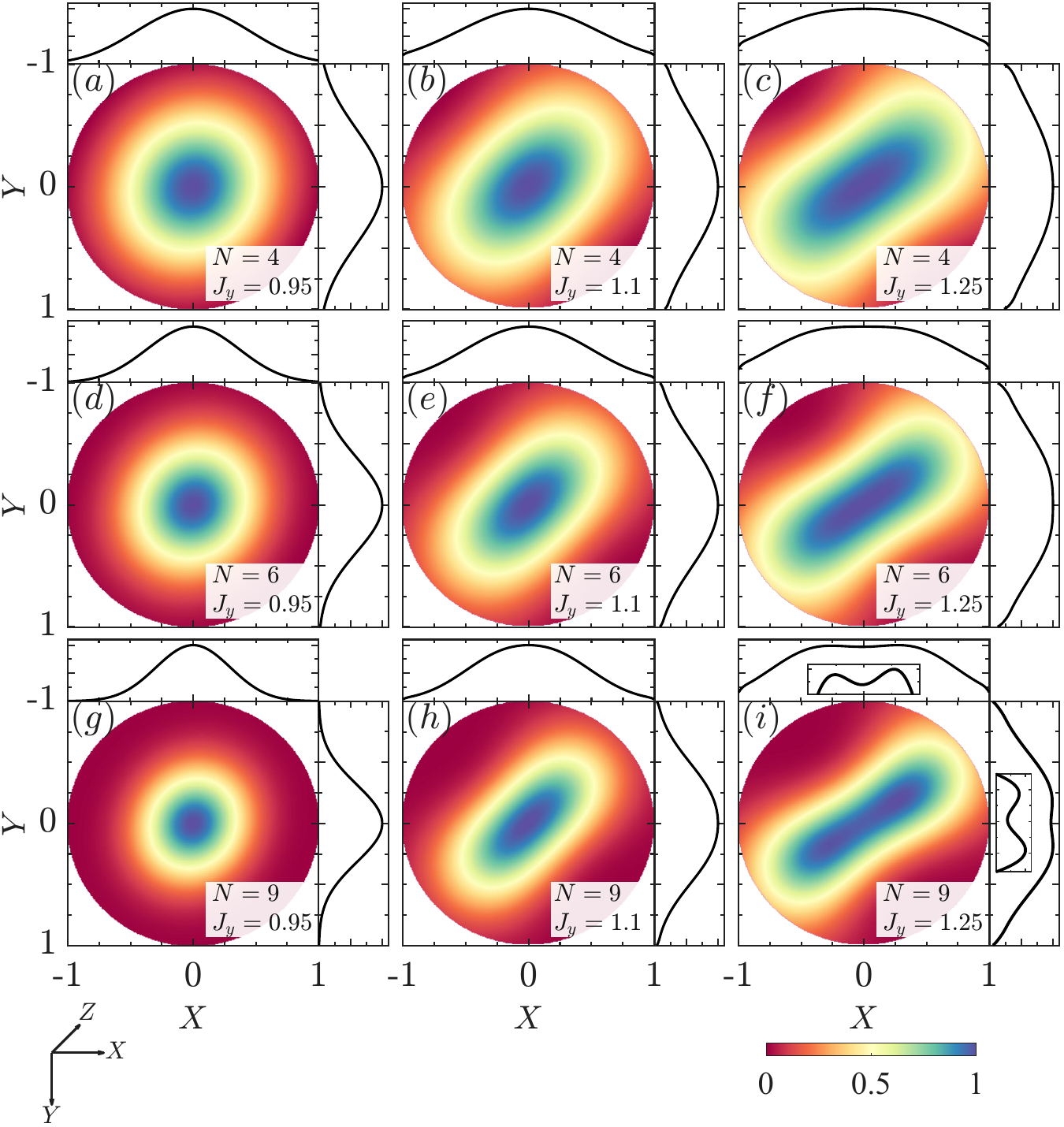}
    \caption{\label{Fig:Husimi}
    Husimi-$Q$ distributions of the steady states viewed from the negative $z$ direction.
    The corresponding parameters are indicated in each panel, and the side profiles show the projections along the $X$ and $Y$ directions.
    With increasing spin number $N$ and coupling strength $J_y$, the Husimi-$Q$ distribution develops an asymmetric double-peak structure, as shown in panel (i).
    Here, spin numbers $N=N_{x}\times N_{y}$ correspond to $2\times2$, $2\times3$, and $3\times3$ lattice geometries, respectively.
    Unless specified otherwise, $J_x = 0.9$ and $J_z = 1.0$.
    }
\end{figure}

Figure~\ref{Fig:Husimi} shows the steady-state Husimi-$Q$ distributions. 
For small $J_y$, the distributions exhibit a single dominant maximum with an approximately symmetric profile, indicating that the steady state is concentrated in one semiclassical phase-space region. 
Increasing $J_y$ stretches the distribution along a diagonal direction, revealing a deformation associated with the underlying $\mathbb{Z}_2$ symmetry. 
This deformation becomes more pronounced for larger system sizes. 
For the largest-size case shown in panel (i), the elongated distribution develops a visible tendency toward an asymmetric double-peak structure, suggesting the emergence of two competing semiclassical sectors. 
The unequal weights of these incipient peaks should be understood as a finite-size precursor of symmetry-sector selection rather than strict spontaneous symmetry breaking at finite size. 
Thus, the Husimi-$Q$ distribution provides semiclassical phase-space evidence for the onset of symmetry-sector formation, motivating the trajectory-level and ensemble-level symmetry diagnostics below.

\begin{figure}[!htpb]
    \centering
    \includegraphics[width=0.48\textwidth]{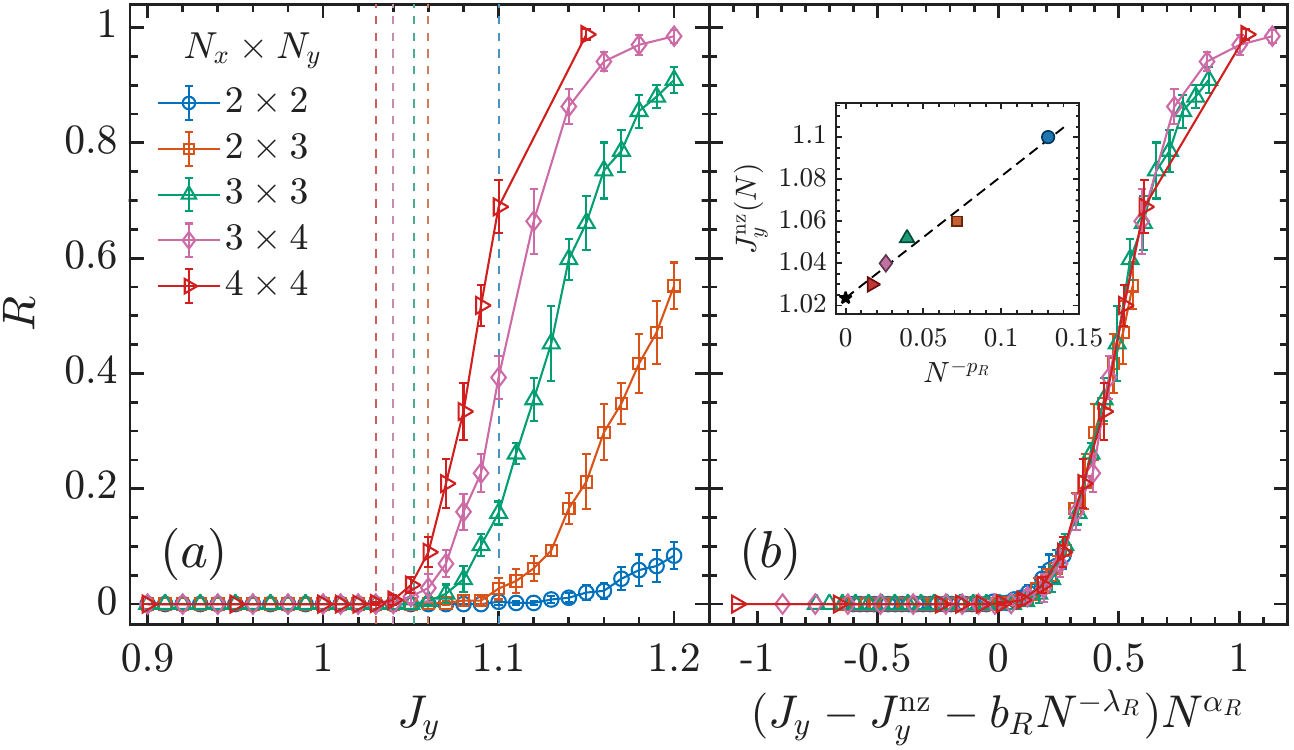}
    \caption{\label{Fig:QT}
    Trajectory-resolved symmetry-breaking ratio $R$.
    For each point, the ratio $R$ is evaluated at $\gamma t=120$ as the fraction of symmetry-violating trajectories among 100 realizations.
    Markers and error bars denote the mean and standard deviation over 10 independent groups, with the threshold set to $\epsilon=10^{-10}$.
    (a) Ratio $R$ versus $J_y$ for various system sizes; dashed lines mark the non-zero (nz) points $J_y^{\text{nz}}(N)$, at which $R$ first becomes nonzero.
    (b) Finite-size scaling collapse of the data in (a). Inset: scaling fit of $J_y^{\text{nz}}(N)$ and its extrapolation to the thermodynamic limit ($N\to \infty$).}
\end{figure}

\blueit{Strong-symmetry breaking.--} 
We unravel Eq.~(\ref{eq:lindblad}) into quantum trajectories using the Monte Carlo wave-function method (MCWF), where each infinitesimal time interval $\delta t$ is resolved into either a deterministic no-jump evolution or a stochastic quantum jump~\cite{Breuer2010,Dalibard1992,Molmer1993}.
The no-jump branch is governed by the non-Hermitian Hamiltonian $H_{\rm NH}=H-\frac{i}{2}\sum_{j=1}^{N}\sigma_j^+\sigma_j^-$, which preserves the symmetry as $[H_{\rm NH},U]=0$.
The jump branch comprises local decay jumps generated by $\sigma_j^-$, which occur with probability $dp_j=dt\langle\psi(t)|\sigma_j^+\sigma_j^-|\psi(t)\rangle$, and transform covariantly as $U\sigma_j^-U^\dagger=-\sigma_j^-$.
Therefore, the trajectory dynamics consists of symmetry-preserving no-jump evolution interrupted by discrete symmetry-covariant jumps, without introducing continuous stochastic fields that could explicitly bias a symmetry sector.
This makes the MCWF particularly suitable for resolving trajectory-level symmetry breaking.

Following the trajectory-level strong-symmetry condition defined above, a quantum state $\rho_\alpha$ preserves the symmetry if $U\rho_\alpha=e^{i\phi_\alpha}\rho_\alpha$.
Thus, we quantify the departure from this symmetry condition by the residual
$\Delta_\alpha=\min_{\phi_\alpha}\|U\rho_\alpha-e^{i\phi_\alpha}\rho_\alpha\|_{F}$,
where $\|\cdot\|_{F}$ denotes the Frobenius norm. 
A vanishing residual indicates that the trajectory remains in a definite symmetry sector, whereas a finite residual signals a violation of the symmetry condition.
To quantify the degree of ensemble-level strong-symmetry breaking, we consider an ensemble of $N_{\rm traj}$ quantum trajectories and count the fraction of trajectories that violate the symmetry condition.
The trajectory-resolved symmetry-breaking ratio is then defined as
\begin{equation}
R = \frac{1}{N_{\rm traj}}\sum_{\alpha=1}^{N_{\rm traj}}\Theta(\Delta_\alpha-\epsilon),
\label{eq:R}
\end{equation}
where $\Theta(x)$ is the Heaviside step function and $\epsilon$ is a numerical threshold.
$R=0$ indicates that all trajectories remain symmetry preserving, while a finite $R$ signals the emergence of symmetry-violating trajectories.
To account for the stochasticity of quantum jumps, we repeat the sampling over 10 independent groups of trajectories; markers and error bars in Fig.~\ref{Fig:QT}(a) denote the corresponding mean and standard deviation of $R$, with each group containing $N_{\rm traj}=100$ trajectories.
For small $J_y$, $R$ remains close to zero, indicating that almost all sampled conditioned states satisfy the trajectory-level strong-symmetry condition.
As $J_y$ is increased, $R$ rises rapidly, signaling the onset of trajectory-level strong-symmetry breaking.
The crossover sharpens with increasing system size, suggesting that the symmetry response becomes singular in the thermodynamic limit.

To quantify this onset, we define the finite-size transition point $J_y^{\rm nz}$ as the first value of $J_y$ at which $R$ becomes nonzero.
The extracted $J_y^{\rm nz}(N)$ values are marked by the dashed vertical lines in Fig.~\ref{Fig:QT}(a).
Due to finite-size effects, the onset points shift systematically with spin numbers.
We fit this finite-size shift using the empirical algebraic form
$J_y^{\rm nz}(N)=J_y^{\rm c}+a_{R}N^{-p_{R}}$, as shown in the inset of Fig.~\ref{Fig:QT}(b).
This fit yields $a_{R}\approx0.578$, $p_{R}\approx1.471$, and an estimated thermodynamic critical point $J_y^{\rm nz}\approx1.024$ for strong-symmetry breaking.
To further verify the scaling behavior, we perform a finite-size scaling collapse of the strong symmetry-breaking ratio.
Considering the extrapolated critical point $J_y^{\rm nz}$ and the finite-size correction obtained above, we use the following scaling form:
\begin{equation}
R(J_y,N) \approx F\left[(J_y-J_y^{\rm nz}-b_{R}N^{-\lambda_{R}})N^{\alpha_{R}}\right],
\label{eq:collapse}
\end{equation}
where $F$ is a universal scaling function, $b_{R}\approx 1.477$ and $\lambda_{R}\approx 2.072$ are the finite-size correction parameters obtained from the fit [see Supplemental Material~\cite{supp} for the collapse without correction].
$\alpha_{R}\approx 0.770$ is a critical exponent optimized to achieve the best collapse.
Figure~\ref{Fig:QT}(b) shows that data for different system sizes collapse onto a single universal curve when plotted in terms of the scaling variable in Eq.~(\ref{eq:collapse}).
The excellent collapse demonstrates that the trajectory-level symmetry-breaking ratio obeys a single-parameter scaling form and captures the critical behavior of the dissipative phase transition.

\blueit{Weak-symmetry breaking.--}
The trajectory-level symmetry breaking discussed above reveals a qualitative reorganization of the conditioned steady-state ensemble.
However, such a reorganization alone does not establish a dissipative phase transition, since the latter is ultimately defined at the level of the ensemble steady state in the thermodynamic limit.
For any finite system, the steady-state density matrix remains compatible with the weak symmetry of $\Ls$ and therefore does not exhibit a strict symmetry breaking.
\begin{figure}[!htpb]
    \centering
    \includegraphics[width=0.48\textwidth]{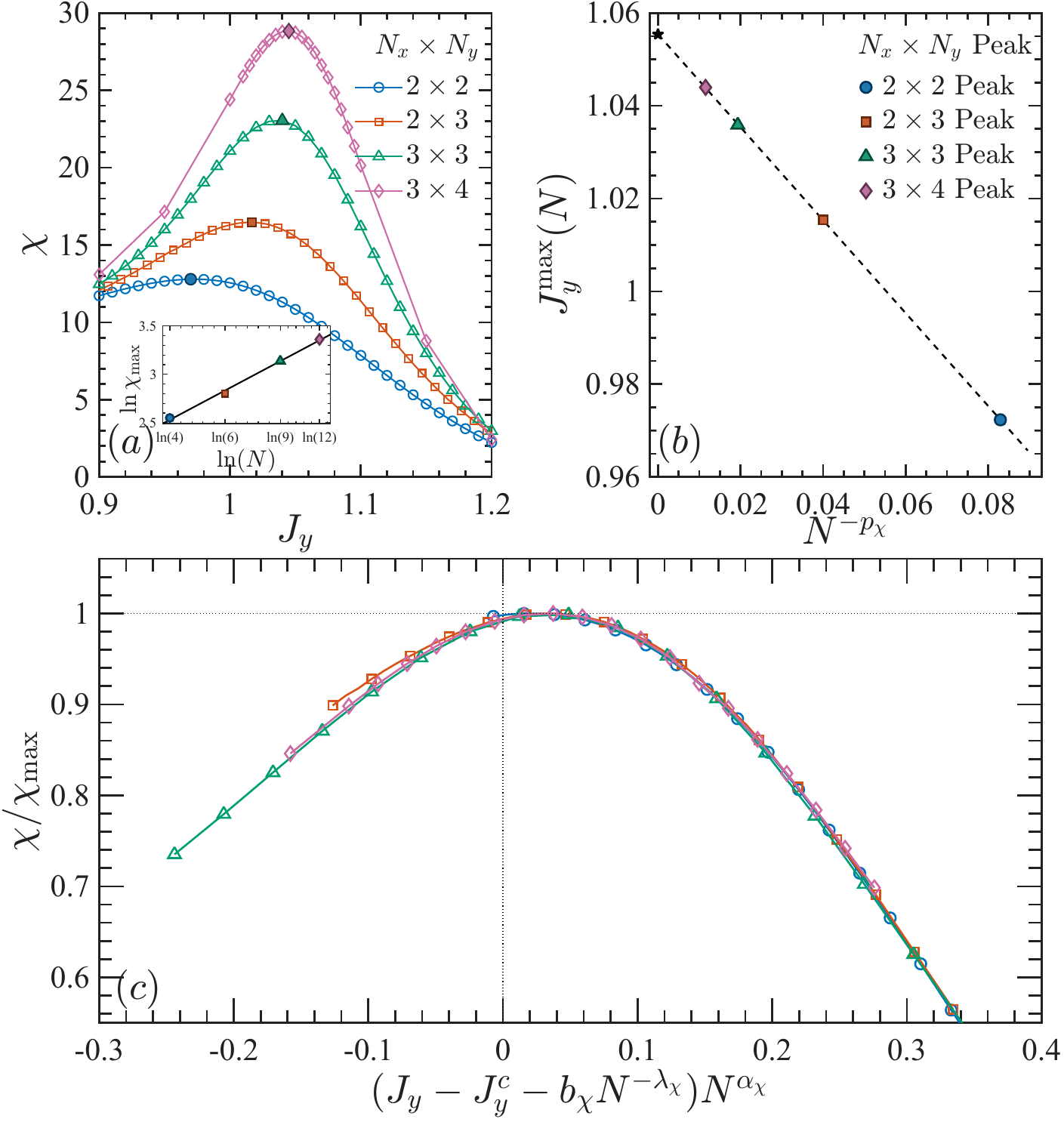}
    \caption{\label{Fig:sstrace}
    Weak-symmetry parameter sensitivity of the steady state to an infinitesimal symmetry rotation. 
    (a) The sensitivity $\chi$ versus $J_y$ for various system sizes $N$, where solid markers indicate the peaks. 
    Inset: finite-size scaling of these peaks.
    (b) Extrapolation of the peaks to determine the thermodynamic-limit critical point $J_y^c$. 
    (c) Finite-size scaling collapse of the sensitivity curves shown in (a).
    }
\end{figure}
Nevertheless, once the trajectory ensemble develops a pronounced symmetry-sector structure, this structure can leave detectable traces in the finite-size steady state, even though the exact weak symmetry remains preserved. 
To reveal such traces at the ensemble level, we introduce the perturbed symmetry operator $U_\eta=e^{i\eta\sum_j\sigma_j^z}U$, where $\eta$ parametrizes an infinitesimal detuning from the $\mathbb{Z}_2$ symmetry.
We then define the weak-symmetry deviation 
$
\Ds(J_y,N)=
\lim_{\eta\to0^+}
\frac{
\left\|
U_\eta\rho_{\rm ss}(J_y,N)U_\eta^\dagger-\rho_{\rm ss}(J_y,N)
\right\|_1
}{\eta}
$, where $\|\cdot\|_1$ is the trace norm and `ss' denotes the steady state.
This quantity measures the change in the finite-size steady state per unit detuning of the symmetry operation.
As the strength $J_y$ is varied across the dissipative transition, the finite-size steady state is continuously reshaped. 
While $\Ds(J_y,N)$ measures the steady state's response to symmetry detuning at a fixed $J_y$, the phase transition is associated with the sharpest change in this response, reflecting the most rapid reorganization of the steady-state structure near criticality. 
We therefore introduce the parameter sensitivity
\begin{equation}
\chi(J_y,N)=
\frac{\partial \Ds(J_y,N)}{\partial J_y},
\label{eq}
\end{equation}
whose peak identifies the sharpest variation of the weak-symmetry deviation and provides a finite-size estimate of the dissipative critical point.

\begin{figure}[!htpb]
    \centering
    \includegraphics[width=0.48\textwidth]{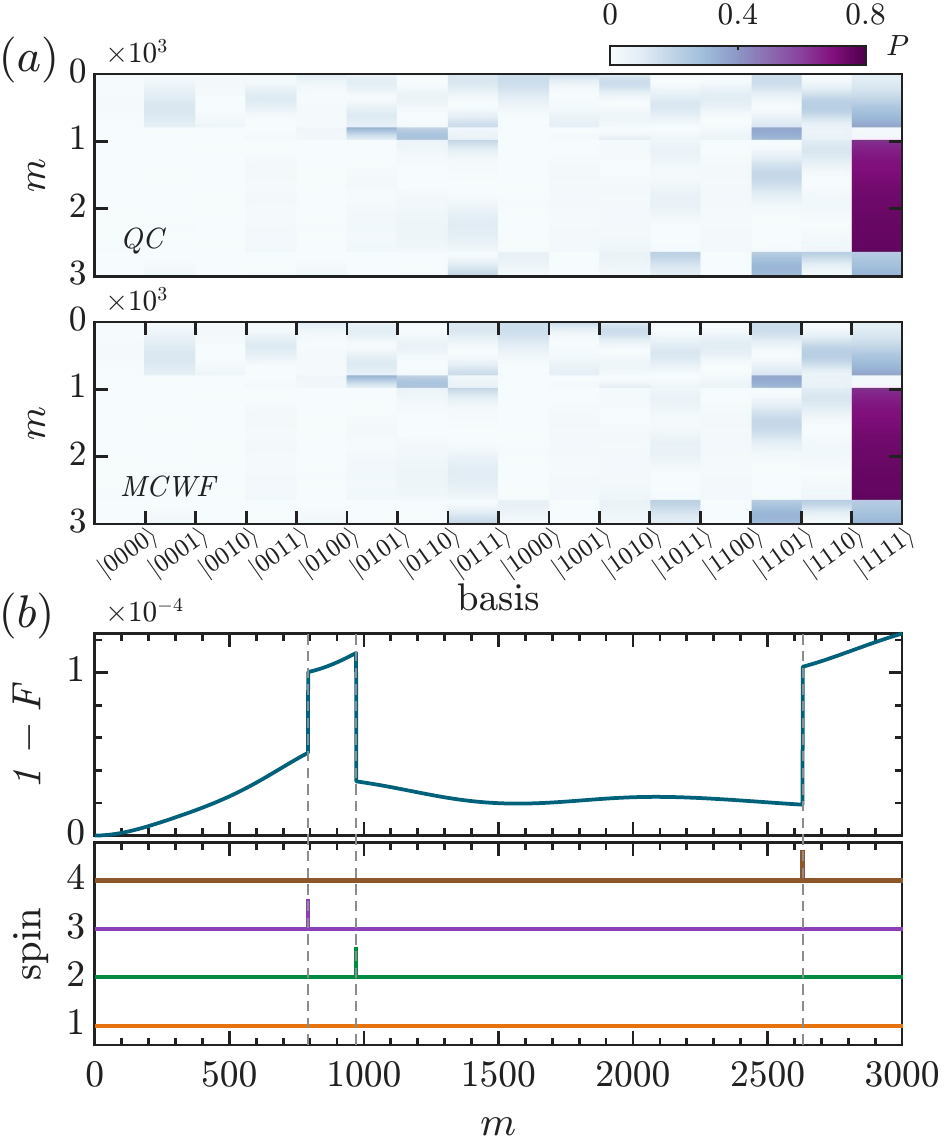}
    \caption{\label{Fig:qiskit} 
    Quantum-circuit validation of the MCWF trajectory for the dissipative spin system ($N=4$).
    (a) Stacked heatmaps comparing the QC (top) and MCWF (bottom)  probabilities of computational basis states as a function of the evolution step under the same quantum-jump sequence. 
    (b) Upper panel: the state infidelity $1-F$ between the QC and MCWF trajectories, where $F=|\langle\psi_{\rm QC}|\psi_{\rm MCWF}\rangle|^2$.
    Lower panel: binary jump records for the four spins, with 0 indicating no jump and 1 indicating a decay event.
    All panels use $m=3000$ evolution steps.
    }
\end{figure}

Figure~\ref{Fig:sstrace}(a) shows the sensitivity $\chi$ as a function of $J_y$ for different system sizes, with the peak positions $J_y^{\max}(N)$ marked by solid symbols.
For the smallest lattice, $\chi$ varies smoothly and exhibits only a weak maximum.
As the number of spins increases, this maximum is strongly enhanced and evolves into a pronounced, sharper peak, indicating that the steady state becomes increasingly sensitive to infinitesimal deviations from the exact symmetry operation.
This size-dependent enhancement is consistent with the emergence of a singular sensitivity in the thermodynamic limit.
The inset of Fig.~\ref{Fig:sstrace}(a) shows that the peak height grows approximately as a power law, $\chi_{\max}\sim N^{\kappa}$ with $\kappa\approx0.748$, further supporting the critical nature of the sensitivity.
Given this increasingly singular behavior with system size, the peak position $J_y^{\max}(N)$ provides a natural finite-size estimator of the underlying critical point.
Taking into account the finite-size drift of the peak positions, we fit the data in Fig.~\ref{Fig:sstrace}(b) using the algebraic form
$J_y^{\max}(N)=J_y^c+a_{\chi} N^{-p_{\chi}}$.
The fit yields $a_{\chi}\approx-1.002$, $p_{\chi}\approx1.796$, and an estimated critical point $J_y^c\approx1.055$, in good agreement with previous studies~\cite{Lee2013,Jin2016,Rota2017,Rota2018a,Biella2018,Li2022c,Verstraelen2023,Matsumoto2025a}.
The simultaneous growth of $\chi_{\max}$ and drift of $J_y^{\max}(N)$ suggest a universal finite-size scaling structure.
We therefore collapse the full sensitivity curves using the same drift-corrected scaling form as in Eq.~(\ref{eq:collapse}), but with independently fitted parameters,
\begin{equation}
\chi(J_y,N)/\chi_{\max}(N)
\approx
G\left[
\left(J_y-J_y^{c}-b_{\chi}N^{-\lambda_{\chi}}\right)N^{\alpha_{\chi}}
\right],
\label{eq:collapse_chi}
\end{equation}
where $G$ is the universal scaling function.
With $b_{\chi}\approx-0.677$, $\lambda_{\chi}\approx1.438$, and $\alpha_{\chi}\approx0.591$, the data collapse well onto a single curve [Fig.~\ref{Fig:sstrace}(c)], thereby confirming that the weak-symmetry sensitivity faithfully captures the finite-size scaling of the dissipative criticality.
Together, these results show that symmetry-resolved diagnostics reveal the dissipative transition through both trajectory-level sector formation and ensemble-level weak-symmetry sensitivity.

\blueit{Experimental feasibility.--}
The symmetry-resolved framework developed above relies on access to quantum trajectories and steady-state properties of the dissipative spin dynamics.
A central ingredient is therefore the ability to realize the MCWF evolution itself.
Here we propose a quantum-circuit algorithm for simulating the MCWF dynamics, whose detailed construction is given in the End Matter.

To benchmark the QC algorithm, we perform a trajectory-by-trajectory comparison between the QC implementation and MCWF. 
Importantly, the MCWF trajectory is conditioned on the same sequence of quantum-jump events recorded in the QC simulation through mid-circuit measurements of the ancillary qubits.
This ensures that the QC and MCWF dynamics correspond to the same realization of the stochastic evolution.
Figure~\ref{Fig:qiskit}(a) shows the time evolution of the probabilities in computational basis states obtained from the QC simulation and from the MCWF trajectory. 
The two results agree very well throughout the evolution. 
To further quantify the simulation accuracy, we evaluate the state infidelity between the two trajectories in Fig.~\ref{Fig:qiskit}(b). 
The infidelity remains below $10^{-4}$ throughout the evolution, while its residual fluctuations coincide with the recorded quantum-jump events shown at the bottom of Fig.~\ref{Fig:qiskit}(b), confirming the faithful reproduction of the MCWF trajectory by the circuit algorithm.
Since both the symmetry diagnostics and steady-state properties can be extracted from quantum trajectories, the results demonstrate the experimental feasibility of the symmetry-resolved framework.

\blueit{Conclusions.--} In summary, we have developed a symmetry-resolved framework for characterizing dissipative criticality in open quantum systems. 
By combining Husimi-$Q$ phase-space analysis, trajectory-level diagnostics, and the weak-symmetry sensitivity of the steady state, we reveal how symmetry breaking emerges across different levels of description. 
The trajectory dynamics exhibits a clear onset of strong symmetry breaking, while the weak-symmetry sensitivity displays critical scaling behavior and provides an accurate estimate of the dissipative critical point. 
We propose a quantum-circuit algorithm that faithfully reproduces the underlying MCWF dynamics, offering a feasible route toward experimental realization~\cite{Sander2025a}. 
Our results establish symmetry as an operational probe of dissipative phase transitions and provide a general framework for exploring symmetry-resolved critical phenomena in open quantum many-body systems. 
More broadly, the enhanced sensitivity associated with symmetry breaking and critical fluctuations may offer opportunities for dissipative quantum sensing and quantum metrology~\cite{Mattes2025,Garbe2020,Fernandez-Lorenzo2017}, where symmetry-resolved observables could serve as powerful resources for detecting weak perturbations and probing nonequilibrium many-body dynamics.

\blueit{Acknowledgments.--}
This work was supported by the National Natural Science Foundation of China under Grant No. 12605045, No. 12605041, No. 92565105, No. 12204395, the fundamental research funds for the central universities (Grant No. GK202601011).
We acknowledge financial support from the Quantum Science and Technology-National
Science and Technology Major Project (2024ZD0300600), Hong Kong RGC No. 14301425, No. 24308323, and No. C4050-23GF, the Space Application System of China
Manned Space Program, and CUHK Direct Grant.
W. L. acknowledges support from the National Natural Science Foundation of China (Grant No. 12304389), the Scientific Research Foundation of NEU (Grant No. 01270021920501*115).
J. J. was supported by the Natural Science Foundation of Liaoning Province No. 2025-MS-009.
J. P. acknowledges support from the Postdoctoral Fellowship Program of CPSF under Grant No. 2025M773387.
X. L. also acknowledges support from the National Natural Science Foundation of China (Grant No. 12247101), the Fundamental Research Funds for the Central Universities (Grant No. lzujbky-2025-jdzx07), the Natural Science Foundation of Gansu Province (22JR5RA389, No.25JRRA799), and the ‘111 Center’ under Grant No. B20063.

\blueit{Data availability.--} Data supporting this article are openly available~\cite{li_2026_21058436}.

\section{End Matter}

\blueit{Quantum-circuit algorithm.---}
We now introduce the QC algorithm used to simulate the MCWF dynamics underlying the dissipative spin model.
The key idea is to replace the stochastic jump sampling in MCWF by mid-circuit measurements of ancillary qubits.
Each circuit step therefore corresponds to one MCWF time step, in which the state is conditionally updated through either the no-jump branch or the jump branch.

\begin{figure}[!htpb]
    \centering
    \includegraphics[width=0.48\textwidth]{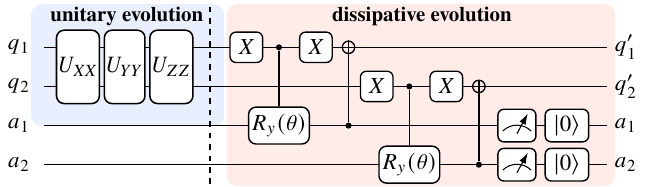}
    \caption{\label{Fig:QC}
    Quantum circuit for simulating one MCWF quantum-jump update of the dissipative spin system, illustrated for the minimal interacting unit with two system qubits, i.e., $q_{1}$ and $q_{2}$.
    The circuit implements the MCWF step by separating the unitary interaction evolution from the dissipative update, where each system qubit is coupled to an ancilla qubit that is measured and reset to record the jump event.
    The parameters used in Fig.~\ref{Fig:qiskit} are $\delta t = 0.001$ and $J_y = 1.2$.
    }
\end{figure}

Figure~\ref{Fig:QC} displays the circuit layout for a minimal two-qubit interacting unit. The extension to larger systems is straightforward by assigning one ancilla qubit to each dissipative channel.
The circuit is partitioned into an initial unitary-evolution block and a subsequent dissipative-update block.

The unitary-evolution block implements a first-order Trotter decomposition of the coherent dynamics,
\begin{equation}
e^{-iH\delta t}
\approx
U_{XX}U_{YY}U_{ZZ},
\end{equation}
with
\begin{equation}
U_{\mu\mu}
=
e^{-iJ_\mu\delta t\,\sigma_j^\mu\sigma_k^\mu},
\qquad
\mu=x,y,z.
\end{equation}
Equivalently, in terms of the two-qubit rotations, the gate angles are
\begin{equation}
\theta_{XX}=2J_x\delta t,\quad
\theta_{YY}=2J_y\delta t,\quad
\theta_{ZZ}=2J_z\delta t.
\end{equation}

Then, the dissipative part is implemented by coupling each system qubit $q_j$ to an ancilla qubit $a_j$.
Consider a system qubit in the state
$
|\psi\rangle=c_0|0\rangle+c_1|1\rangle,
$
while the ancilla is initialized in
$
|0\rangle_{ a}.
$
The two $X$ gates surrounding the controlled-$R_y(\theta)$ operation ensure that the ancilla rotation is applied conditionally on the system state.
After the dissipative block, the joint state becomes
\begin{equation}
|\psi\rangle|0\rangle_{ a}
\rightarrow
\left(c_1|1\rangle
+
c_0\cos\frac{\theta}{2}|0\rangle\right)|0\rangle_{ a}
+
c_0\sin\frac{\theta}{2}|1\rangle|1\rangle_{ a}
.
\end{equation}
Choosing
$
\cos\frac{\theta}{2}
=
\sqrt{1-\delta t}
$, we have $\sin\frac{\theta}{2}=\sqrt{\delta t}$, and therefore the above expression can be rewritten as
\begin{equation}
|\psi\rangle|0\rangle_{ a}
=
\left(
c_1|1\rangle
+
c_0\sqrt{1-\delta t}\,
|0\rangle
\right)
|0\rangle_{ a}
+
c_0\sqrt{\delta t}\,
|1\rangle|1\rangle_{ a}.
\label{eq:jointstate}
\end{equation}
The single-spin construction above applies independently to each dissipative channel.
For the channel associated with spin $j$, Eq.~(\ref{eq:jointstate}) can be recast into the Kraus form. 
Before measuring the ancilla $a_j$, the dissipative block realizes the conditional mapping
\begin{equation}
|\psi\rangle |0\rangle_{a_j}
\rightarrow
M_0^{(j)}|\psi\rangle |0\rangle_{a_j}
+
M_1^{(j)}|\psi\rangle |1\rangle_{a_j},
\end{equation}
where, to first order in \(\delta t\),
\begin{equation}
M_0^{(j)}
=
1-\frac{\delta t}{2}\sigma_j^+\sigma_j^-,
\qquad
M_1^{(j)}
=
\sqrt{\delta t}\,\sigma_j^- .
\end{equation}

The ancilla measurement selects the corresponding MCWF branch.
If the measurement gives $a_j=0$, no jump occurs and the system state is updated as
\begin{equation}
|\psi\rangle
\rightarrow
\frac{M_0^{(j)}|\psi\rangle}
{\sqrt{
\langle\psi|
M_0^{(j)\dagger}M_0^{(j)}
|\psi\rangle
}},
\end{equation}
which is the no-jump update.
If the measurement gives \(a_j=1\), a jump occurs on spin \(j\), and the system state is updated as
\begin{equation}
|\psi\rangle
\rightarrow
\frac{M_1^{(j)}|\psi\rangle}
{\sqrt{
\langle\psi|
M_1^{(j)\dagger}M_1^{(j)}
|\psi\rangle
}}
=
\frac{\sigma_j^-|\psi\rangle}
{\sqrt{
\langle\psi|
\sigma_j^+\sigma_j^-
|\psi\rangle
}},
\end{equation}
which is exactly the MCWF jump update.
After the measurement, the ancillas are reset to $|0\rangle_{a_j}$ and reused in the next time step.
Repeating this procedure generates a sequence of ancilla measurement outcomes,
\begin{equation}
\{a_j(\delta t),a_j(2\delta t),\cdots\},
\end{equation}
which is precisely the quantum-jump record of the MCWF trajectory.
Therefore, each circuit time step has a one-to-one correspondence with one MCWF update:
\begin{equation}
a_j=0
\Longleftrightarrow
\mathrm{no\ jump},
\qquad
a_j=1
\Longleftrightarrow
\mathrm{jump\ on\ spin}\ j .
\end{equation}

\bibliography{filterRef}

\end{document}


\beginsupplement

\title{Supplemental Material for
\mbox{``Symmetry-Resolved Criticality in Dissipative Quantum Phase Transitions''}}

\author{Xingli Li\orcidlink{0000-0003-2339-6557}}
\affiliation{Department of Physics, The Chinese University of Hong Kong, Shatin, New Territories, Hong Kong, China}
\affiliation{The Chinese University of Hong Kong Shenzhen Research Institute, Shenzhen 518057, China}
\affiliation{Lanzhou Center for Theoretical Physics, Key Laboratory of Theoretical Physics of Gansu Province, Key Laboratory of Quantum Theory and Applications of MoE, Gansu Provincial Research Center for Basic Disciplines of Quantum Physics, Lanzhou University, Lanzhou 730000, China}

\author{Yan Li\orcidlink{0009-0007-7218-133X}}
\email{yanliphy@snnu.edu.cn}
\affiliation{School of Physics and Information Technology, Shaanxi Normal University, Xi’an, 710062, P. R. China}

\author{Jinghui Pi\orcidlink{0000-0002-6899-8132}}
\affiliation{The Chinese University of Hong Kong Shenzhen Research Institute, Shenzhen 518057, China}

\author{Wenlin Li\orcidlink{0000-0003-2032-0530}}
\affiliation{College of Sciences, Northeastern University, Shenyang 110819, China}
\affiliation{Research Center for Quantum Physics, Northeastern University, Shenyang 110819, China}

\author{\mbox{Shuai Geng}\orcidlink{0009-0000-8687-0493}}
\affiliation{School of Physics, Dalian University of Technology, Dalian 116024, China}

\author{Lin Shang\orcidlink{0009-0007-5840-2351}}
\affiliation{School of Physics, Dalian University of Technology, Dalian 116024, China}

\author{Jiasen Jin\orcidlink{0000-0002-7118-7488}}
\email{jsjin@dlut.edu.cn}
\affiliation{School of Physics, Dalian University of Technology, Dalian 116024, China}

\author{Yangqian Yan\orcidlink{0000-0002-3237-5945}}
\email{yqyan@cuhk.edu.hk}
\affiliation{Department of Physics, The Chinese University of Hong Kong, Shatin, New Territories, Hong Kong, China}
\affiliation{State Key Laboratory of Quantum Information Technologies and Materials, The Chinese University of Hong Kong, Hong Kong SAR, China}
\affiliation{The Chinese University of Hong Kong Shenzhen Research Institute, Shenzhen 518057, China}

\date{\today}

\maketitle

\section{Scaling collapse without finite-size correction}

To further examine the robustness of the finite-size scaling analysis, we perform an additional scaling collapse without introducing a finite-size correction to the critical point.

\begin{figure}[!htpb]
    \centering
    \includegraphics[width=0.8\textwidth]{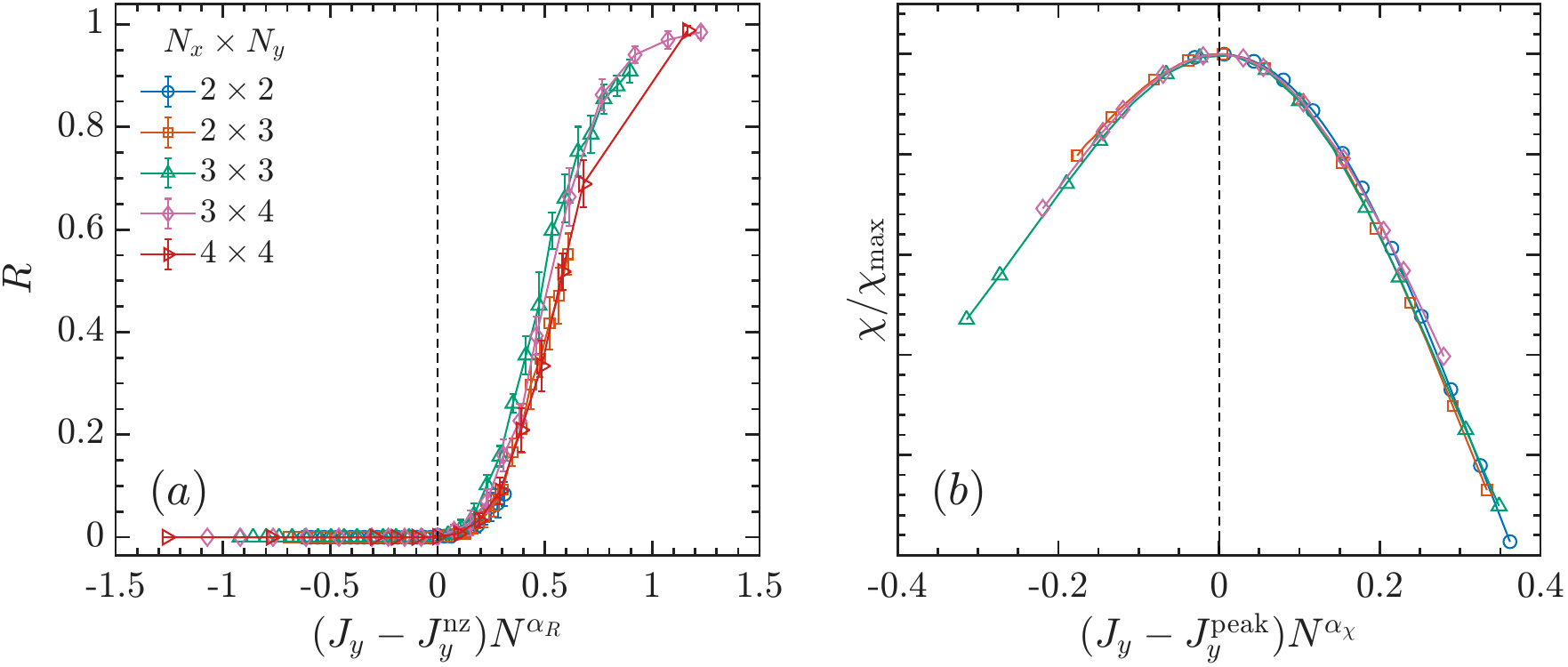}
    \caption{\label{Fig:raw}
    Finite-size scaling collapse without the finite-size correction.
    (a) Collapse of the strong-symmetry breaking ratio $R$ as a function of $(J_y-J_y^{\rm nz})N^{\alpha_R}$, where $J_y^{\rm nz}$ is the finite-size onset point of nonzero $R$.
    (b) Peak-centered collapse of the weak-symmetry sensitivity, plotted as $\chi/\chi_{\max}$ versus $(J_y-J_y^{\rm peak})N^{\alpha_{\chi}}$, where $J_y^{\rm peak}$ and $\chi_{\max}$ denote the peak position and peak height for each system size.
}
\end{figure}

In the main text, the scaling collapse was carried out using a corrected scaling variable that accounts for the finite-size drift of the peak positions.
Here, we set this correction to zero and directly use the extrapolated nonzero point and critical point obtained from the thermodynamic-limit extrapolation.
For the trajectory-resolved symmetry-breaking ratio $R$, the uncorrected scaling form is given by
\begin{equation}
R(J_y,N) \approx F\left[(J_y-J_y^{\rm nz})N^{\alpha_{R}}\right],
\end{equation}
where $J_y^{\rm nz}\approx1.024$ is the same thermodynamic-limit extrapolated nonzero point used in the main text. 
For the weak-symmetry parameter sensitivity $\chi$, we similarly collapse the data according to
\begin{equation}
\chi(J_y,N)/\chi_{\max}(N)
\approx
G\left[
\left(J_y-J_y^{c}\right)N^{\alpha_{\chi}}
\right],
\end{equation}
where $J_y^c\approx1.055$ is the same thermodynamic-limit extrapolated critical point obtained from the peak positions of $\chi$ and used in the main text. 
In both cases, only the finite-size correction term is omitted in the present collapse.

Figure~\ref{Fig:raw}(a) shows the scaling collapse of the ratio $R$ using the uncorrected scaling function, with $\alpha_{R}\approx0.8196$.
This value is slightly different from the corrected result $\alpha_R\approx0.770$ reported in the main text, reflecting the absence of the finite-size correction term.
Nevertheless, the data collapse reasonably well around the onset region, indicating that the trajectory-resolved symmetry-breaking behavior is robust against the removal of the finite-size correction.
Figure~\ref{Fig:raw}(b) shows the corresponding scaling collapse of the weak-symmetry parameter sensitivity $\chi$ using the uncorrected scaling variable, with $\alpha_{\chi}\approx 0.6467$. 
Similar to the case of the ratio $R$, the fitted exponent exhibits only a small shift compared with the corrected value used in the main text, while the curves for different system sizes still collapse well near the critical region. 
These results confirm that omitting the finite-size correction does not affect the identification of the transition points or the associated scaling behavior.

\section{Results for periodic boundary conditions}
In the main text, we focused on the results obtained under open boundary conditions (OBC). 
To further verify that the observed symmetry-resolved critical behavior is not tied to a particular choice of boundary condition, we also perform the same analysis under periodic boundary conditions (PBC).
In this section, we present the PBC results for the steady-state Husimi-$Q$ distribution, the trajectory-resolved symmetry-breaking ratio and the weak-symmetry parameter sensitivity.

\begin{figure}[!htpb]
    \centering
    \includegraphics[width=0.5\textwidth]{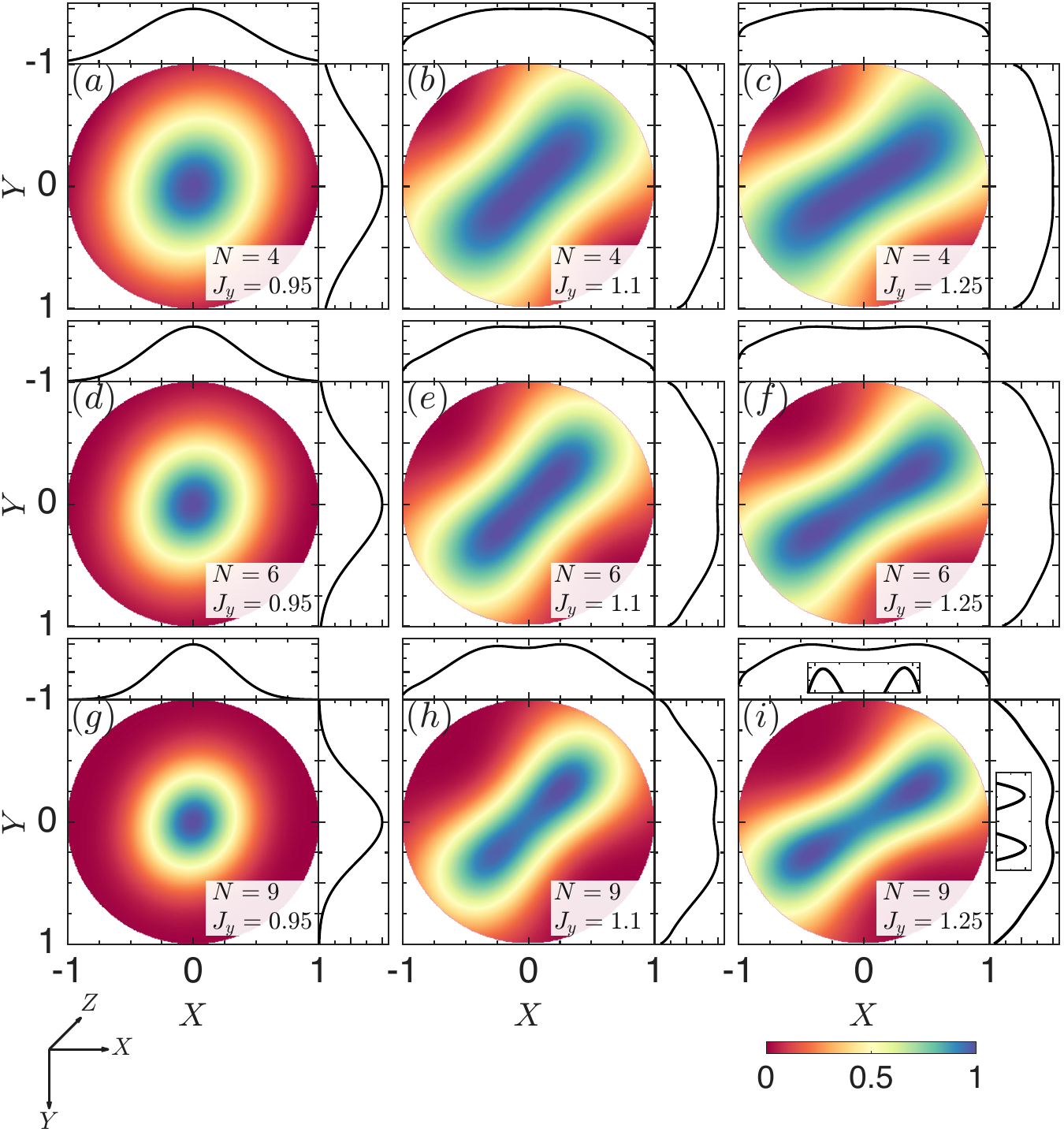}
    \caption{\label{Fig:Husimi_PBC} Husimi-$Q$ distributions of steady states under PBC.
    The corresponding parameters are indicated in each panel, and the side profiles show the projections along the $X$ and $Y$ directions.
    With increasing spin number $N$ and coupling strength $J_y$, the Husimi-$Q$ distribution develops an asymmetric double-peak structure, as shown in panel (i).
    Here, spin numbers $N=N_{x}\times N_{y}$ correspond to $2\times2$, $2\times3$, and $3\times3$ lattice geometries, respectively.
    The remaining parameters are fixed at $J_x = 0.9$ and $J_z = 1.0$.
}
\end{figure}

In Fig.~\ref{Fig:Husimi_PBC}, we show the Husimi-$Q$ distributions of the steady states under PBC, with all other parameters the same as those used in the main text. 
The overall evolution of the phase-space structure is similar to the OBC case. 
For small $J_y$, the distribution is dominated by a single maximum, while increasing $J_y$ leads to an elongation of the distribution along a diagonal direction. 
This deformation becomes more pronounced for larger system sizes, reflecting the same tendency toward symmetry-sector formation as discussed in the main text.

\begin{figure}[!htpb]
    \centering
    \includegraphics[width=0.75\textwidth]{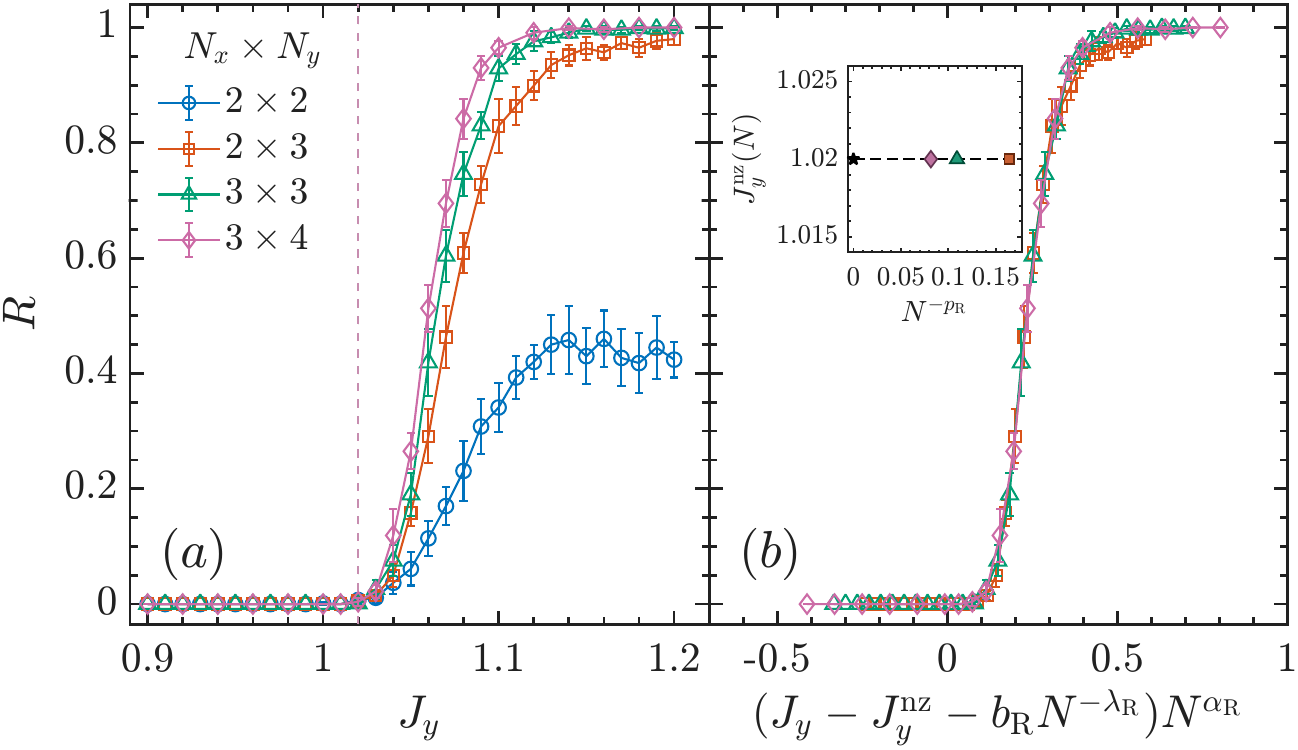}
    \caption{\label{Fig:QT_PBC}
    Trajectory-resolved symmetry-breaking ratio $R$ under PBC.
    For each point, the ratio $R$ is evaluated at $\gamma t=120$ as the fraction of symmetry-violating trajectories among 100 realizations.
    Markers and error bars denote the mean and standard deviation over 10 independent groups, with the threshold set to $\epsilon=10^{-10}$.
    (a) Ratio $R$ versus $J_y$ for various system sizes; dashed lines mark the nonzero points $J_y^{\mathrm{nz}}(N)$, where the onset of $R>0$ occurs.
    (b) Finite-size scaling collapse of the data in (a). Inset: scaling fit of $J_y^{\mathrm{nz}}(N)$ and its extrapolation to the thermodynamic limit ($N\to\infty$).
}
\end{figure}

In the largest-size case shown in panel $(i)$, the distribution develops a clear double-peak structure. 
Compared with the OBC result, the two peaks under PBC are more balanced in height, with a peak-height ratio of approximately $99.8\%$.
This near balance indicates that the two semiclassical sectors are almost equally weighted under PBC, while the appearance of the double-peak structure still provides phase-space evidence for the onset of symmetry-sector formation.

Figure~\ref{Fig:QT_PBC}(a) presents the results of the trajectory-resolved symmetry-breaking ratio $R$ under PBC. 
For all system sizes except $N=4$, $R$ increases with $J_y$ and eventually approaches the saturated value $R=1$, indicating the onset of trajectory-level symmetry breaking under PBC. 
Under PBC, boundary-induced inhomogeneity is absent and all sites share the same local environment. 
Once the system is large enough to accommodate the relevant trajectory-level symmetry-broken configurations, the finite-size dependence of $R$ becomes weak, which explains the close agreement among the $N=6$, $9$, and $12$ results.

The exceptional behavior of the $N=4$ curve is attributed to the small-cluster character of the $2\times2$ geometry. 
The trajectory-resolved ratio $R$ is a finite-time diagnostic rather than an infinite-time steady-state order parameter, so the observation time should be chosen within the metastable window. 
We use a fixed observation time $\gamma t=120$ for all system sizes. 
While varying the observation time within the accessible window does not change the qualitative finite-size trend for $N\geq6$, the $N=4$ ring shows a much faster decay of $R$. 
This indicates that its trajectory-level symmetry-broken metastable regime has a lifetime comparable to the observation time.
Moreover, the $N=4$ size is too small to represent a bulk-like periodic configuration, since fluctuations can wrap around the entire ring and symmetry-related trajectory sectors are strongly mixed. 
We therefore regard $N=4$ as a small-size effect and do not use it to infer the scaling behavior.
Excluding this smallest cluster, the $N=6$, $9$, and $12$ data exhibit a clear scaling collapse, as shown in Fig.~\ref{Fig:QT_PBC}(b).
For consistency with the main text, the same corrected scaling form is adopted. 
The resulting collapse indicates that the finite-size scaling behavior of the trajectory-resolved ratio $R$ is preserved under PBC. 
The corrected scaling collapse yields an extrapolated nonzero point $J_{y}^{\rm nz}\approx 1.02$, which is very close to the OBC value ($J_{y}^{\rm nz}\approx 1.024$) reported in the main text. 
The remaining fitting parameters are $b_{R}\approx -0.145$, $\lambda_{R}\approx 0.8359$, and $\alpha_{R}\approx 0.5630$. 
This agreement further confirms that the trajectory-level onset point is robust against the change of boundary condition.

\begin{figure}[!htpb]
    \centering
    \includegraphics[width=0.9\textwidth]{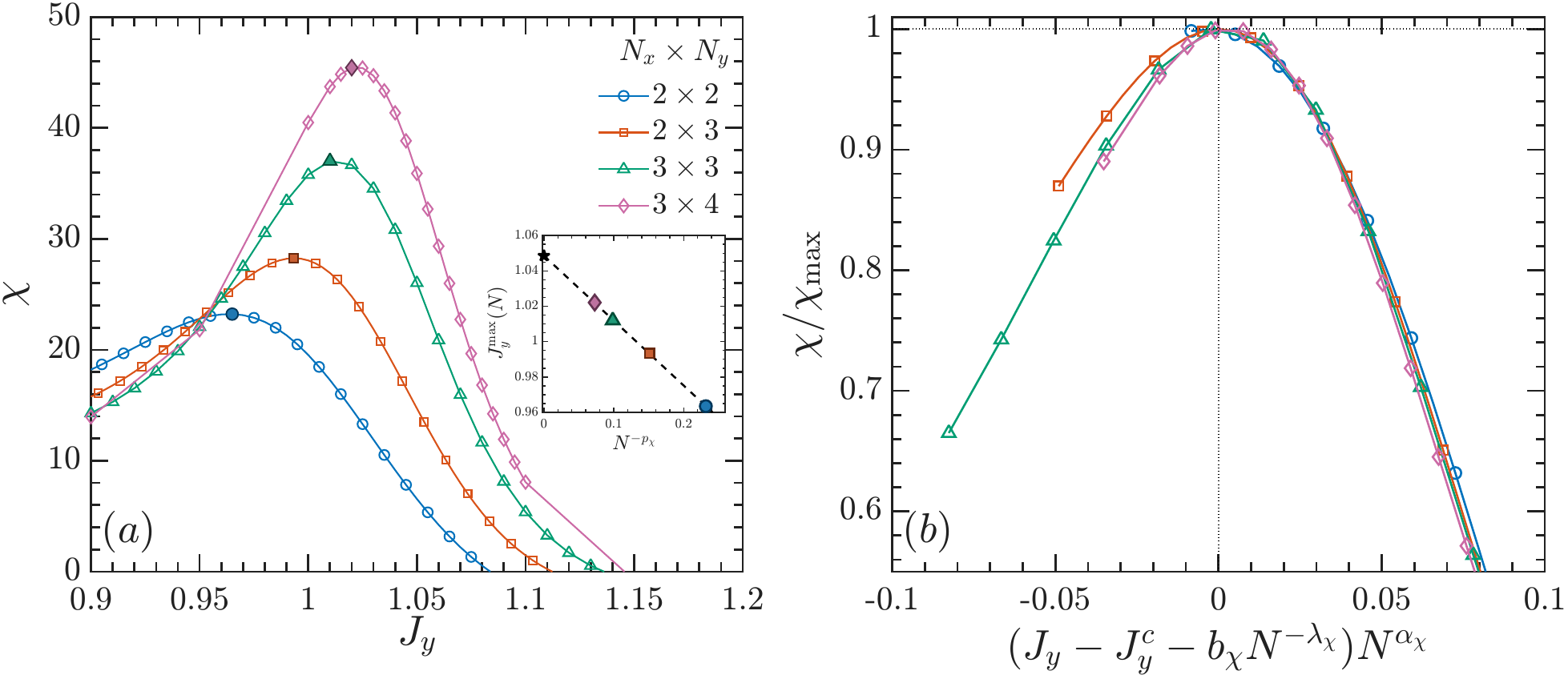}
    \caption{\label{Fig:sstrace_PBC}
    Weak-symmetry parameter sensitivity of the steady state under PBC.
    (a) The sensitivity $\chi$ versus $J_y$ for various system sizes $N$, where solid markers indicate the peaks. Inset: finite-size extrapolation of the peak positions used to determine the thermodynamic-limit critical point $J_y^c$.
    (b) Finite-size scaling collapse of the sensitivity curves shown in (a).
}
\end{figure}

We next turn to the weak-symmetry parameter sensitivity of the steady state under PBC. 
Figure~\ref{Fig:sstrace_PBC}(a) shows the corresponding results for different system sizes, with the peak positions marked by solid symbols. 
Similar to the OBC results in the main text, $\chi$ develops a pronounced peak as the system size increases. 
The enhancement and sharpening of this peak indicate that the steady state becomes increasingly sensitive to infinitesimal deviations from the exact symmetry operation. 
This size-dependent behavior is consistent with the emergence of a singular weak-symmetry response in the thermodynamic limit.
The inset shows the thermodynamic-limit extrapolation of the peak positions using
$J_y^{\max}(N)=J_y^c+a_{\chi} N^{-p_{\chi}}$, from which the critical point under PBC is extracted. 
The fit gives $J_y^c\approx 1.0486$, close to the OBC value reported in the main text. 
This agreement further indicates that the weak-symmetry critical response is robust against the choice of boundary condition.
We further perform the scaling collapse of $\chi$ under PBC using the same corrected scaling function as in the OBC analysis. 
The resulting collapse is shown in Fig.~\ref{Fig:sstrace_PBC}(b), with fitting parameters are $\alpha_{\chi}\approx 0.2159$, $b_{\chi}\approx -0.3213$, and $\lambda_{\chi}\approx 0.9816$. 
The good collapse near the critical region further confirms that the weak-symmetry scaling behavior is preserved under PBC.

Overall, the PBC analysis reproduces the main symmetry-resolved signatures observed under OBC, including the Husimi-$Q$ double-peak tendency, the onset of trajectory-level symmetry breaking, and the size-enhanced weak-symmetry response. 
The close agreement of the extracted transition points further confirms that our conclusions are robust against the choice of boundary condition.



\bibliography{filterRef}